\documentclass[final,12pt]{article}

\usepackage{graphicx}
\usepackage{titling}
\usepackage{float}
\usepackage{subfig}
\usepackage{cite}
\usepackage{authblk}

\title{The TT-PET Data Acquisition and Trigger System}

\author[1]{Y. Bandi}
\author[2]{Y. Favre}
\author[2]{D. Ferrere}
\author[1]{D. Forshaw}
\author[1]{R. H\"anni}
\author[2]{D. Hayakawa}
\author[2]{G. Iacobucci}
\author[1]{P. Lutz}
\author[1]{A. Miucci}
\author[2]{L. Paolozzi}
\author[2]{E. Ripiccini}
\author[1]{C. Tognina}
\author[2]{P. Valerio}
\author[1]{M. Weber}
\affil[1]{University of Bern, Sidlerstrasse 5, CH-3012 Bern, Switzerland}
\affil[2]{University of Geneva, Rue du Général-Dufour 24, Geneva, Switzerland}

\begin{document}
\maketitle
\begin{abstract}

This paper describes the data acquisition and trigger system of the Thin Time-of-flight PET (TT-PET) scanner. The system is designed to read out in the order of 1000 pixel sensors used in the scanner and to provide a reference timing signal to each sensor in order to measure tie differences of better than 30 ps. This clock distribution is measured to have a jitter of less than 4 ps at the sensors. Collected data is locally processed before being forwarded to storage. Data flow as well as control, configuration and monitoring aspects are are also addressed. 
\end{abstract}


\setcounter{tocdepth}{2}
\tableofcontents
\thispagestyle{empty}
\cleardoublepage


\section{Introduction}\label{S:1}
 
\subsection{The TT-PET Project}\label{subsec:project}
The TT-PET collaboration is developing a compact time-of-flight PET scanner for small animals which is designed to be inserted into existing commercial MRIs machines such as the nanoScan 3T MRI \cite{mediso:3t}.
Traditional PET scanners use detector modules composed of scintillating crystals coupled to photomulipliers or photodiodes using light guides.
However, the TT-PET scanner will use a series fully monolithic silicon pixel detectors based on Silicon\,-\,Germanium (SiGe) BiCMOS technology arranged in a multi-layered structure. The design of the scanner will profit from the silicon detectors experience in the context of high-energy physics experiments as the ATLAS detector at the LHC \cite{Aad:2008zzm}.

\subsection{Scanner Overview}\label{subsec:scanner_over}
The scanner and its supporting mechanics is designed to be housed inside the removable RF-coil of the nanoScan 3T MRI machine \cite{mediso:3t}. While all the readout electronics and power supplies will be located outside of the MRI machine, and connected using long flat shielded kapton cables. This design strategy is typical in high energy physics and has the advantage of removing sensitive electronics, such as FPGA's from the alternating magnetic fields inside the MRI machine, making shielding and cooling simpler.\\

\begin{figure}[!htb] 
    \centering
    \includegraphics[width=0.80\textwidth]{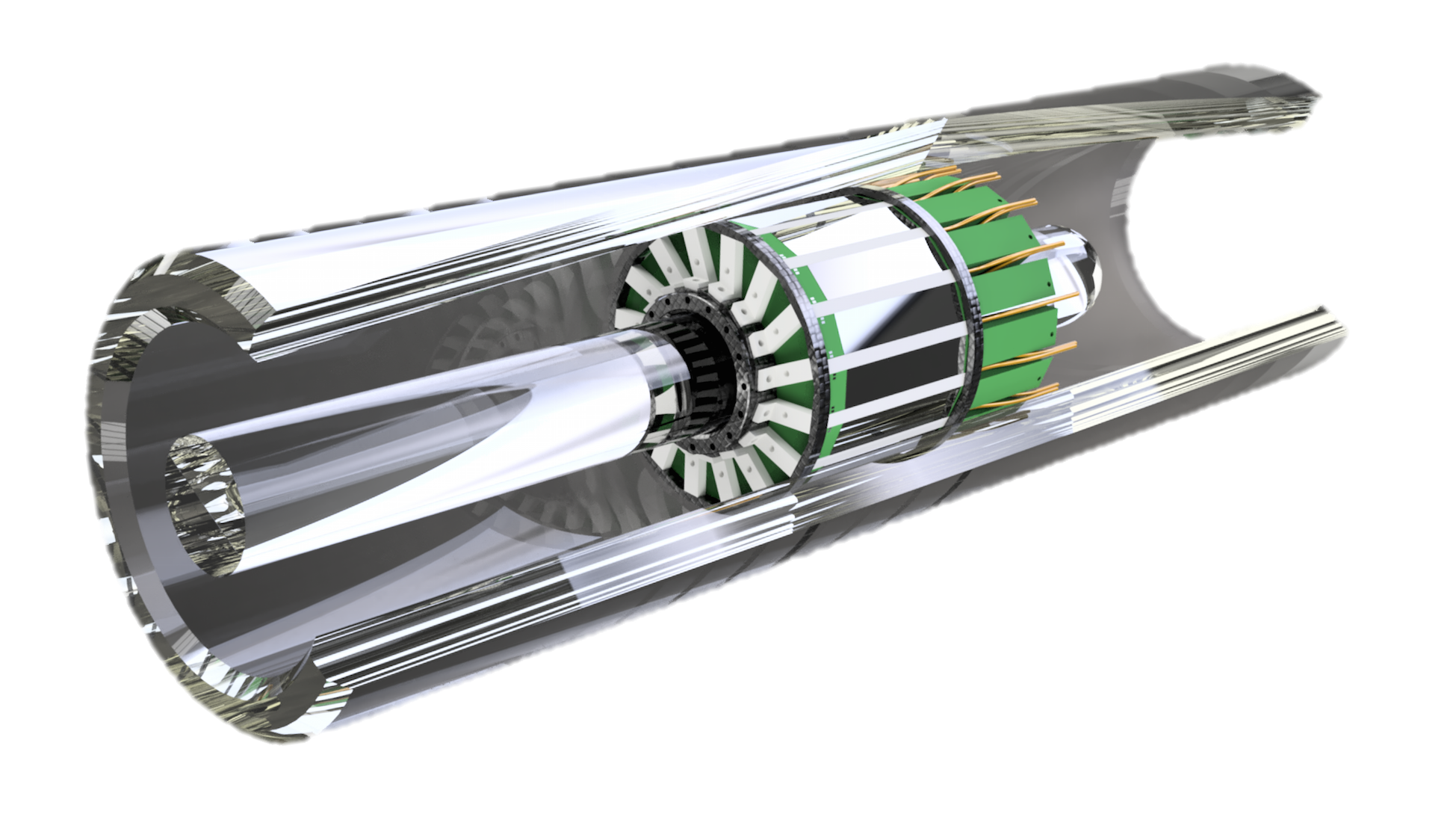}
    \caption{3D CAD model rendering of the TT-PET scanner housed inside the removable RF-coil of a nanoscan 3T (clear) and with a small animal insert (clear) placed in its center. Cooling and electrical services continue outside the RF-coil but are not shown.}
    \label{fig:scanner}
\end{figure}

The sensitive part of the scanner is composed of 16 detector modules, referred to as tower modules, and 16 plastic/ceramic cooling blocks arranged in an alternating pattern.
The novel sensors \cite{ttpet:p0} used for the modules have been measured to have excellent efficiency and a timing resolution of around 100 ps in beam tests. \cite{ttpet:demo}
A single layer ring structure is formed and can be seen in (Figure\,\ref{fig:scanner}). The inner and outer diameters of the scanner ring are 33\,mm and 79\,mm respectively, including internal and external carbon fiber support rings; the inner radius of the RF-coil is 80\,mm. A single tower module is 15.3\,mm tall (Figure\,\ref{fig:tower}) and is composed of 60 single detection layers which are stacked directly on top of each other, creating a superimposed sampling (sandwich) structure.\\

The design of the scanner has been optimised using Monte Carlo simulations performed using Geant4 \cite{ALLISON2016186} and FLUKA\,\cite{fluka}. More details about the scanner layout are available at \cite{mechanics_paper}.\\


In order to have a high detection efficiency, the lead foil placed in-between the sensor layers is used as a converter material. This converts the gamma-rays produced in the imaging sample to electrons in the pixel sensors.
The expected data rate for the scanner and towers modules was simulated and used to guide the development of the readout DAQ system. A point-like F18 source, with an activity of 50\,MBq was simulated in the center of a cyclindrical water volume with R=18\,mm, to simulate the presence of a small animal with a radioactive tracer roughly 10x stronger than what would normally be used. A single photon detection rate of 19.2\,MHz is expected assuming a minimum energy deposition of 20\,keV inside a sensor. While for a coincidence, two single hits have to be detected in a time window of less than 6.25\,ns $(\,|\,t1\,-\,t2\,|\,<\,6.25\,$ns$\,)$ which corresponds to the L1 clock frequency of the daq system. The expected data rate for all coincidence candidate events is 10\,MHz. However, since the hits have to occur in separate detection towers and the resulting Line of Response (LOR) must intersect with the water cylinder. A coincidence hit rate of $\sim$\,2.4\,MHz or 1.2\,MHz per tower is expected, which is sent to the daq computer for data storage before post-processing.
A selective triggering scheme using both hardware and software is needed, where each event is associated to a unique clock cycle ID (ccID) and time stamp with multiple stages of data storage and suppression to remove unwanted background events, enabling real time reconstruction and image processing.
This is equivalent to a total data flow rate of 1.7\,Gbit/s from the readout system to the storage computer.
More details about the data-rate simulation can be found at \cite{simulations}
The front-end (monolithic sensors) electronics send event data, via shielded kapton flex cables, sub-divided into super modules (section\,\ref{subsec:supermodule}) to their relevant Tower Control boards. The Tower Control boards organise the event data and associate them with a clock cycle identifier (ccID). At the start of a daq scan all bits of the ccID are set to 0. \\

The Tower Control boards store the event data in temporary storage while the ccID data fragments are sent to the Central Trigger Processor (CTP) for comparison. If the ccID values for the recorded events pass a first stage of coincidence trigger, the CTP subsequently requests the full event data to be sent along the full daq chain to the CTP. Currently, the data is formatted as follows: start bit $+$ 4\,bit header $+$ 10 bit ccID $+$ data from chip $+$ stop bit. The 10\,bit ccID is composed of a 9\,bit eventID $+$ 1\,bit sensor ID value. The "data from chip" is 54\,bits long and contains the event data for a single sensor. In order to preserve data-integrity a 8b/10b encoding is foreseen to be use at this stage\\
    
A full overview of the the DAQ system, consisting of all stages of the scalable hardware and basic readout architecture is given in section\,\ref{S:2}. Section\,\ref{S:3} focuses on the software and triggering for the system. Section\,\ref{S:4} contains an overview of results of performance tests and of observations from data taking. In section\,\ref{S:5} conclusions and an outlook are presented.


\section{Readout System}\label{S:2}
 
\subsection{System Overview}\label{susec:SO}

\begin{figure}[!hbt] 
\centering
\includegraphics[width=0.99\textwidth]{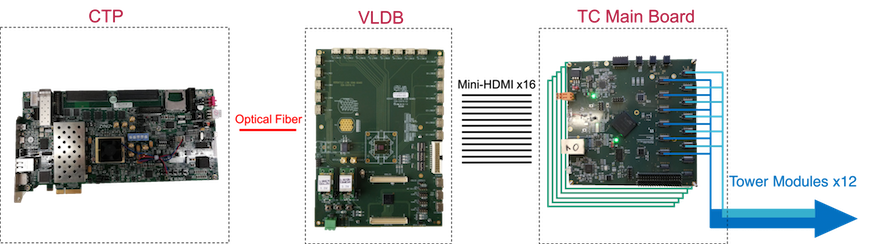}
\caption{Overview of full daq chain showing from left to right; Central Trigger Processor (CTP), Versatile Link Demo Board (VLDB), and TC Main board}
\label{fig:daq-over}
\end{figure}

\begin{figure}[!hbt] 
\centering
\includegraphics[width=0.99\textwidth]{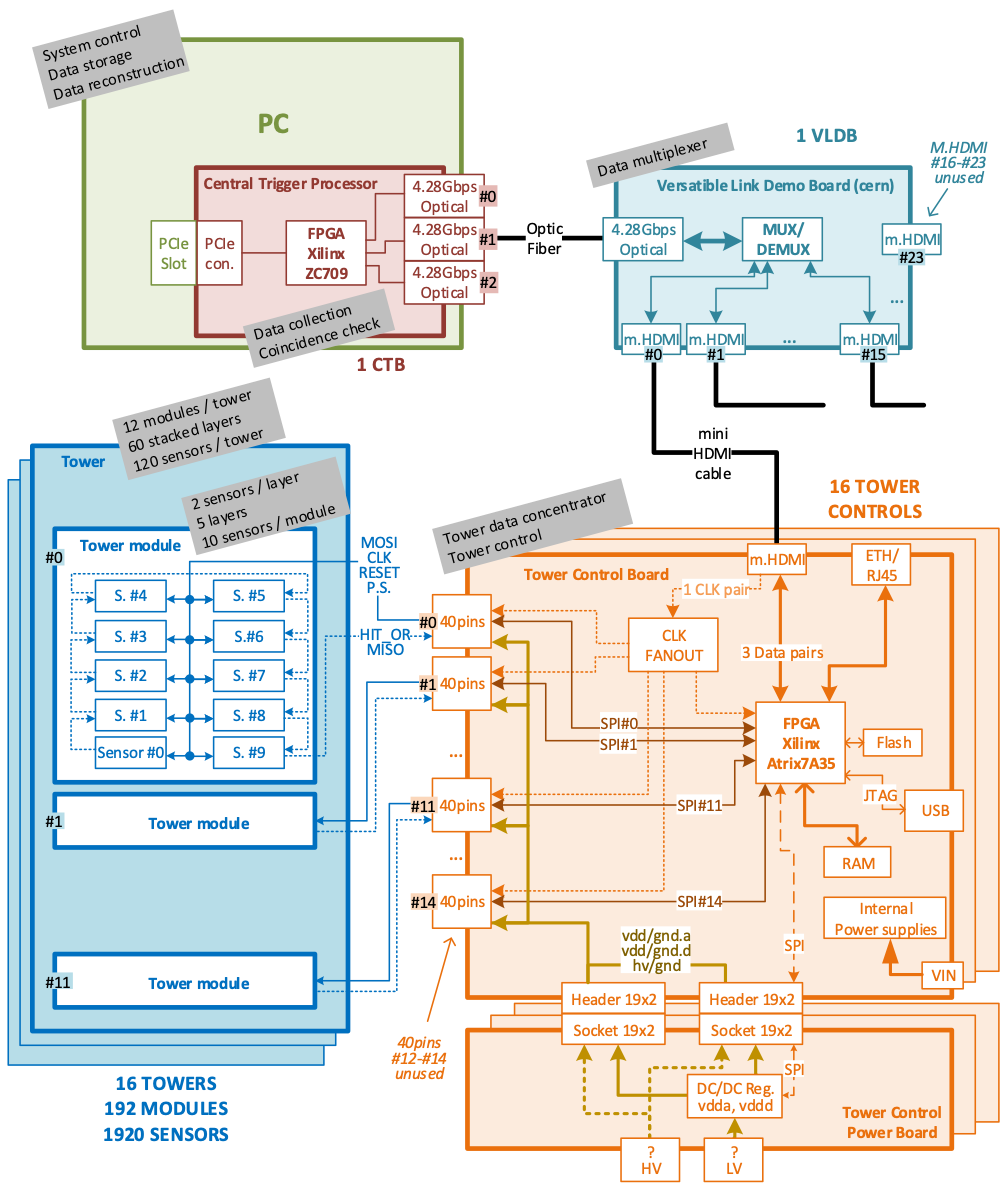}
\caption{Block Diagram of full daq chain showing how each stage is connected together, and by which type of connection.}
\label{fig:daq-STAGE}
\end{figure}

The Data Acquisition (daq) system has been designed to be scalable and therefore has a modular structure. The full daq chain is split into 3 stages, and is shown in Figure\,\ref{fig:daq-over}. Additionally, Figure\,\ref{fig:daq-STAGE} shows a detailed block diagram of the full daq chain with more detailed information.\\

\noindent
\textbf{Stage 1:} Each tower module is connected to a Tower Control (TC) board, which is a custom FPGA board designed to be as small and cost effective as possible. Its core function is to provide temporary data storage\,/\,suppression, 8b10b encoding, high (HV)\,/\,low (LV) operating voltages for the monolithic pixel sensors, and data aggregation before being sent to the second stage of the daq chain. Based on the current scanner geometry 16 TC boards are required, where each board is connected to a tower using multiple thin shielded kapton cables.\\

\noindent
\textbf{Stage 2:} All TC boards are connected to a multiplexer board, named, Versatile Link Demo Board (VLDB) which accepts electrical signals from multiple TC boards and multiplexes their output to a single bitstream before being buffered and converted to an optical signal. The multiplexed data is then sent to third daq stage using a single data link and fiber optic cable at a maximum link speed of 4.28\,Gbps. The current scanner configuration requires only 1 VLDB board and therefore 1 fiber optic cable. however, more boards can be added to the chain as required.\\

\noindent
\textbf{Stage 3:} A large powerful commercial FPGA board (Xilinx VC709) \cite{vldb} referred to as the Central Trigger Processor (CTP) is connected to a control PC using PCIe slots. The CTP board has 4 optical transceiver link ports (4.28\,Gbps) which can be used for communication to the VLDB board/s. The CTP controls the whole daq chain and supplies a low jitter 160\,MHz L0 sync clock to the rest of daq chain. Additional real time processing adds another level of data suppression, and is where coincidence checks are performed. It is foreseen to perform real time image reconstruction using the CTP's powerful FPGA or use additional resources available to the host PC, such as GPU's.\\

Based on the current scanner layout, 16 Tower Control FPGA boards will be connected to a single VLDB, which is then connected using a single 4.28\,Gbps optical fiber to the Central Trigger processor board. However, the daq can easily be up-scaled by simply increasing the number of Tower Control boards which each have 15 electrical input ports (for super-modules) or by adding additional VLDB boards which have 20 e-ports (for Tower Control Boards). The current CTP features 4 optical transceivers, which allows up to 4 VLDB boards to be connected. However, if more optical ports are requires a different version of the CTP can be used which features more optical ports or an intermediate high speed optical switch can be used.\\

\newpage
\subsection{Super-Module}\label{subsec:supermodule}
In order to simplify the scanners readout architecture and services. Each 60 layer tower module is sub-divided into 12 super-modules. This reduces the number of cables and connections that have to be routed outside the RF-Coil to a tower control. Each super-module is composed of 5 consecutively stacked detection layers (2 sensors per layer), whose monolithic sensors are daisy-chained together forming a chain of 10 sensors.\\

To reduce system complexity and easy size constraints for the scanner a linear daisy-chain readout architecture was decided. Due to the harsh environment expected within an MRI machine it has been decided that all data, clock and command lines will be differential. A super module requires 3 global inputs, A synchronous L0 clock, reset line, and command line. The monolithic sensors pass various other signal lines along the linear daisy chain, such as hit data, inter-chip sync clock, and trigger lines. The final sensor on the chain then sends all the trigger and data signals to the TC Main board.\\

\begin{figure}[!hbt] 
\centering
\includegraphics[width=0.99\textwidth]{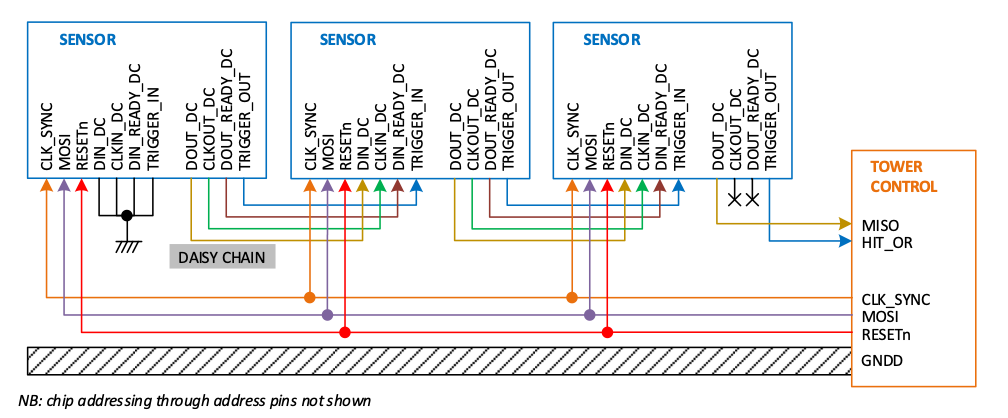}
\caption{Block diagram of a super-module connected to a TC, showing only 3 sensors for simplicity. Common sensor input clock, command (MOSI), data and reset lines are shown going to each sensor in parallel. As well as inter-sensor data, trigger, clock daisy-chain lines. Finally, the last sensor in the chain sends the trigger (HIT\_OR) and output (MISO) to the TC board.}
\label{fig:SUPER-MODULE-BLOCK}
\end{figure}

A block diagram detailing the important electrical connections (excluding all power and ground connections) and their daisy-chaining scheme within a super-module to a TC main board is shown in figure\,\ref{fig:SUPER-MODULE-BLOCK}.\\

\subsection{Tower Control Board}\label{subsec:TC}

Tower Control (TC) boards will be used to control and configure the scanners tower modules and provide all necessary electrical services. The TC has been split into 2 separate boards during development. The main FPGA board (TC Main board) which has all the required functionality for the final system with the exception of providing LV\,/\,HV power to the tower modules. The LV\,/\,HV required will be supplied using an additional board which connects to the TC Main board, which is referred to as the TC Power Board. The TC main board then controls the TC Power board using serial commands (SPI).\\

\begin{figure}[!hbt] 
\centering
\includegraphics[width=0.99\textwidth]{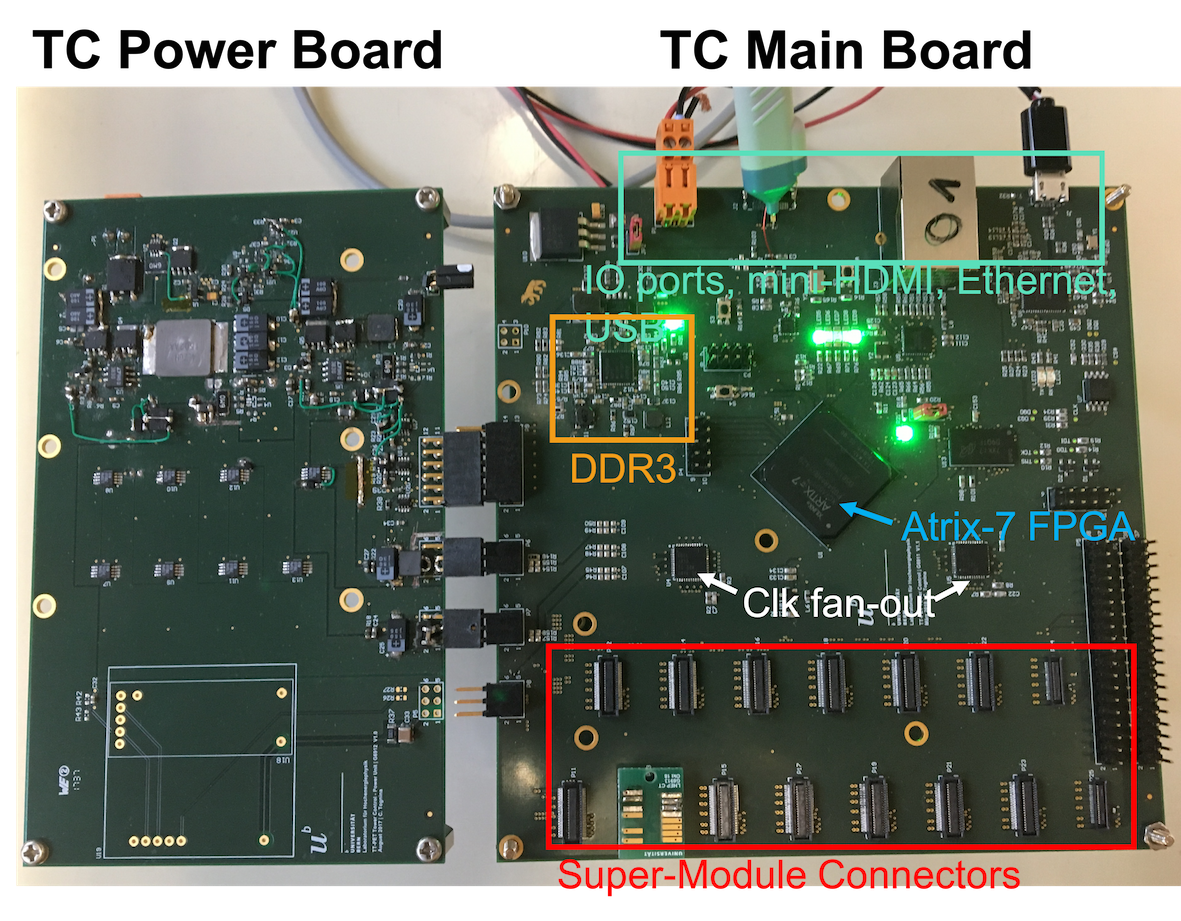}
\caption{Image of a complete Tower Control (TC) development board. Left, TC Power Board and Right, TC main Board. Important features of the board, such as the FPGA (Atrix-7), IO connections, DDR3 ram and super-module connectors are labelled.}
\label{fig:TOWER-CONTROL}
\end{figure}


The TC Main board has been designed to be compact, low cost, and can distribute an external low jitter clock signal via clock fan-outs to each super-module. At its core the main board is composed of an Atrix-7 FPGA, DDR3 memory storage, fast low jitter clock fan-out, and multiple types of IO connectors for to connect to the rest of the daq system and the tower modules. There are 15 molex connectors per board to connect to super-modules; the final connector type is still under investigation. Connection to next stage of the daq (VLDB, section\,\ref{subsec:VLDB}) is done using a single mini-HDMI port. A 160\,MHz synchronous L1 clock is sent to the TC main boards from the VLD. The data rate between the TC and VLDB boards is nominally 160\,Mbps, but can be increased up to 320\,Mbps.\\

 
 The TC power boards will provide all the low and high voltage lines for each tower module requires. The maximum high voltage for the sensors is 300\,V (parallel connection). It is expected that each monolithic sensor draws currents in the nA range. The sensor requires two low voltage lines VDDA and VDDD, these voltage lines are controllable up to 3.3\,V. The expected nominal voltage values are 1.8\,V and 1.2\,V respectively. The TC power board outputs a single channel for HV, VDDA, and VDDD. These are then split and sent to each super-module connector on the TC main board. The table\,\ref{tab:power} shows some key voltages and currents for the TC Power board. The max rated voltages\,/\,currents were calculated assuming a 60 layer tower module where each layer has 2 sensors, making 120 sensors in total.\\

\begin{table}[!hbt]
\begin{center}
\begin{tabular}{|l|c|c|c|c|}
\hline
                    & Minimum & Typical & Maximum &         \\ \hline
Output Voltage      & 1       &         & 3.3     & {[}V{]} \\ \hline
Current Load   VDDA &         &         & 6.7     & {[}A{]} \\ \hline
VDDD                &         &         & 0.7     & {[}A{]} \\ \hline
Input Voltage       & 5.2     & 5.5     & 5.8     & {[}V{]} \\ \hline
Input Current       &         &         & 7.5     & {[}A{]} \\ \hline
\end{tabular}
\end{center}
\label{tab:power}
\end{table}

\subsection{VLDB}\label{subsec:VLDB}

The Versatile Link Demonstrator Board (VLDB) \cite{vldb} is a CERN developed radiation hard optical link evaluation package, which provides magnetic-field-tolerant 4.8\,Gbps data transfer links for communication between front-end (FE) and back-end (BE) electronics. Point-to-point optical links are used to handle Timing, Trigger and Control (TTC) signals with low jitter and deterministic latency, slow control, status monitoring and readout data with GBT-SCA pairs. Each board has 20 electrical link (elink) ports, which offer data rates from 80 to 320\,Mbps. The GBTx also serves as a clock distributor as it recovers the clock from incoming serial data and transmits it to every connected front-end, as well as to 8 phase adjustable clock outputs.\\

\begin{figure}[!hbt] 
\centering
\includegraphics[width=0.95\textwidth]{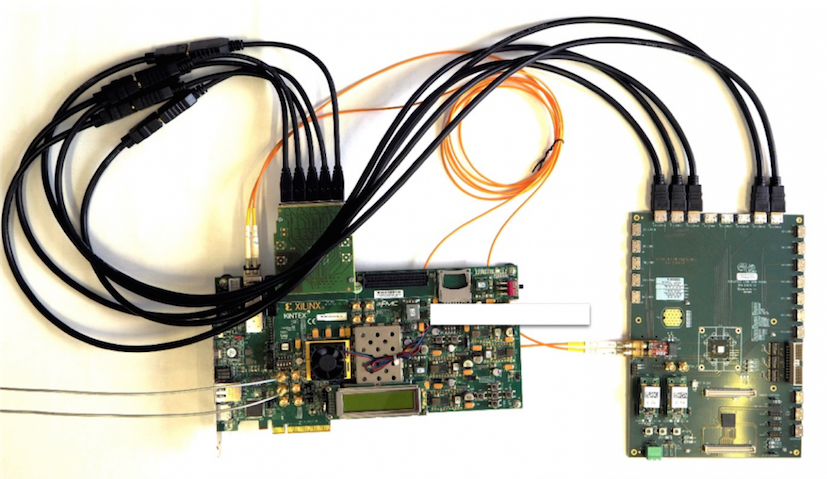}
\caption{VLDB board (right) connected to KC705 evaluation board in loop-back testing configuration. Connection made using elinks over HDMI to an HDMI-FMC adapter \cite{vldb2}.}
\label{fig:vldb1}
\end{figure}

The VLDB will be used as a radiation and magnetic-field-tolerant high bandwidth multiplexer board as part of the TT-PET daq system, which is resistant to SEU's. The onboard GBTx chip will be used to multiplex and de-multiplex command and data signals sent between the daq control PC\,/\,central Trigger Processor (CTP) and each tower module via the tower control boards.\\  

\subsection{Central Trigger Processor}\label{subsec:CTP}

The Central Trigger Processor (CTP) is composed of a single powerful commercial FPGA board. Currently, a Xilinx Virtex-7 FPGA VC709 \cite{vc709} is being used. The CTP will be connected to the daq\,/\,data storage computer via PCIe. The current scanner layout only requires 1 CTP but more CTP boards could be used provided the daq PC has the required free PCIe slots. The CTP is designed to act as the master of the daq chain, requesting interesting data and performing data selection\,/\,triggering (coincidence checks) before the data is sent to the PC and stored for image reconstruction.\\


\section{Data and Trigger}\label{S:3}
 
\subsection{Overview}\label{subsec:trigger-overview}
\begin{figure}[!hbt] 
\centering
\subfloat[][]{\includegraphics[width=0.34\textwidth]{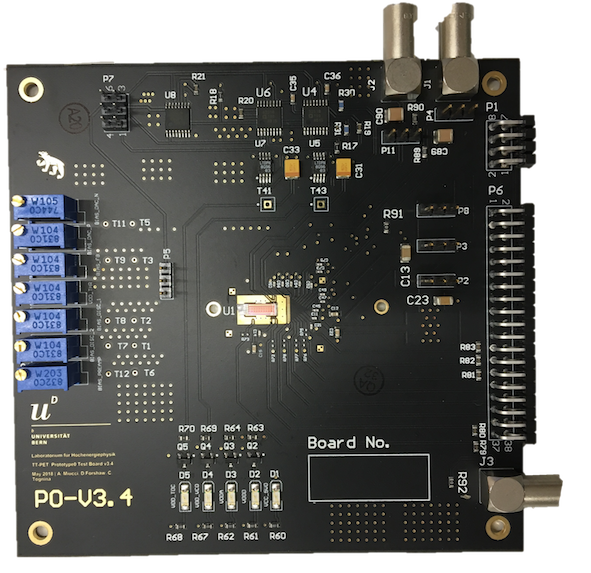}\label{fig:TBV3} }
\subfloat[][]{\includegraphics[width=0.64\textwidth]{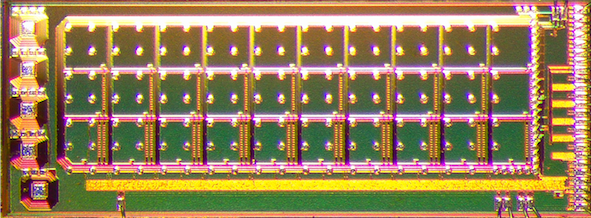}\label{fig:p0} }
\caption{\protect\subref{fig:TBV3} Picture of prototype sensor test PCB (v3). The small prototype sensor is glued to a ground pad at the center of the PCB, with a large hole under the sensor to allow easy passage of charged particles. The sensor is wire bonded (Al 25\,$\mu$m diameter) to the PCB directly. \protect\subref{fig:p0} Top view picture of prototype-0 monolithic pixel sensor\cite{ttpet:p0}.}
\label{fig:TB}
\end{figure}

At the time of writing, only the prototype monolithic sensor \cite{ttpet:p0}  is available for testing with the DAQ system. As a result, a stripped down version of the full DAQ chain was used for sensor characterisation. The TC main board was directly connected to the DAQ PC using USB\,/\,Ethernet and a bespoke version of the TC firmware for readout. Additionally, since the prototype sensor was not designed to be readout via daisy-chain, a test PCB was designed, where each sensor requires its own test PCB and can be seen in Figure\,\ref{fig:TBV3}. The test board contains additional controllable electronics, such as voltage regulators and variable resistors which are controlled via serial communication (SPI).\\

\subsection{Data Flow}\label{subsec:dataflow}

As previously mentioned (section\,\ref{subsec:Bandwidth}) a single event hit rate of 19.2\,MHz is expected for the whole scanner. This translates to a data transmission rate of $\sim$\,1.7\,-\,1.8\,Gbps between the VLDB and the CTP based on the current scanner configuration. This number has been estimated using the expected sensor data protocol (54\,bits) multiplied by the single hit event rate (19.2\,MHz), plus $\sim$\,200\,Mbps for module identification, before being converted with 8b10b encoding. The expected average data rate between each TC and the VLDB is therefore $\sim$\,112\,Mbps per TC, and the average data rate between each super module and TC is $\sim$\,9.3\,Mbps.\\

\subsection{Trigger System}\label{subsec:trig}

The final system is designed to self trigger on gamma ray events produced within the imaging sample. The trigger system is composed of multiple stages of hardware trigger. First, each pixel sensor has a minimum ToT threshold set in each pixel. If the deposited charge is above threshold a trigger command is sent to the TC, and the TC requests and stores the data in a buffer. Next, the TC generates unique ccID values which are tied to the L1 clock of the system. The CTP compares the 10\,bit ccID values for coincidences by checking the ccID values. If the ccID values are within 2 clock cycles the event data is combined into a coincidence event and sent to the CTP for the final trigger stage. The CTP compares the locations of the hit events within the scanners geometry and rejects coincidence events which occurr in the same or neighbouring tower modules. Coincidence events that pass this criteria are saved on the storage PC.\\

\subsection{Synchronisation}\label{subsec:sync}

The VLDB will generate a low jitter 160\,MHz L1 sync clock which is sent to every TC main board in parallel. A low jitter clock fan-out ($\sim$\,fs) on every TC main boards then distributes the L1 clock to every super-module. The L1 clock is sent to every sensor in parallel within a super module. however, since a daisy-chain data readout scheme is used each chip also outputs its clock to the next sensor in the chain to enable time-walk corrections.\\


\section{Performance Tests}\label{S:4}
 

\subsection{Full Readout Chain Emulation}\label{fds}

A logical test bench emulation of the TC main board and Central Trigger Processor (CTP) was performed to demonstrate the full readout chain based on the specifications of the final monolithic sensor. A single TC main board was connected to a CTP; A VLDB board was not included in the simulation, because only 1 TC main board was simulated.\\

\begin{figure}[H] 
\centering
\includegraphics[width=0.99\textwidth]{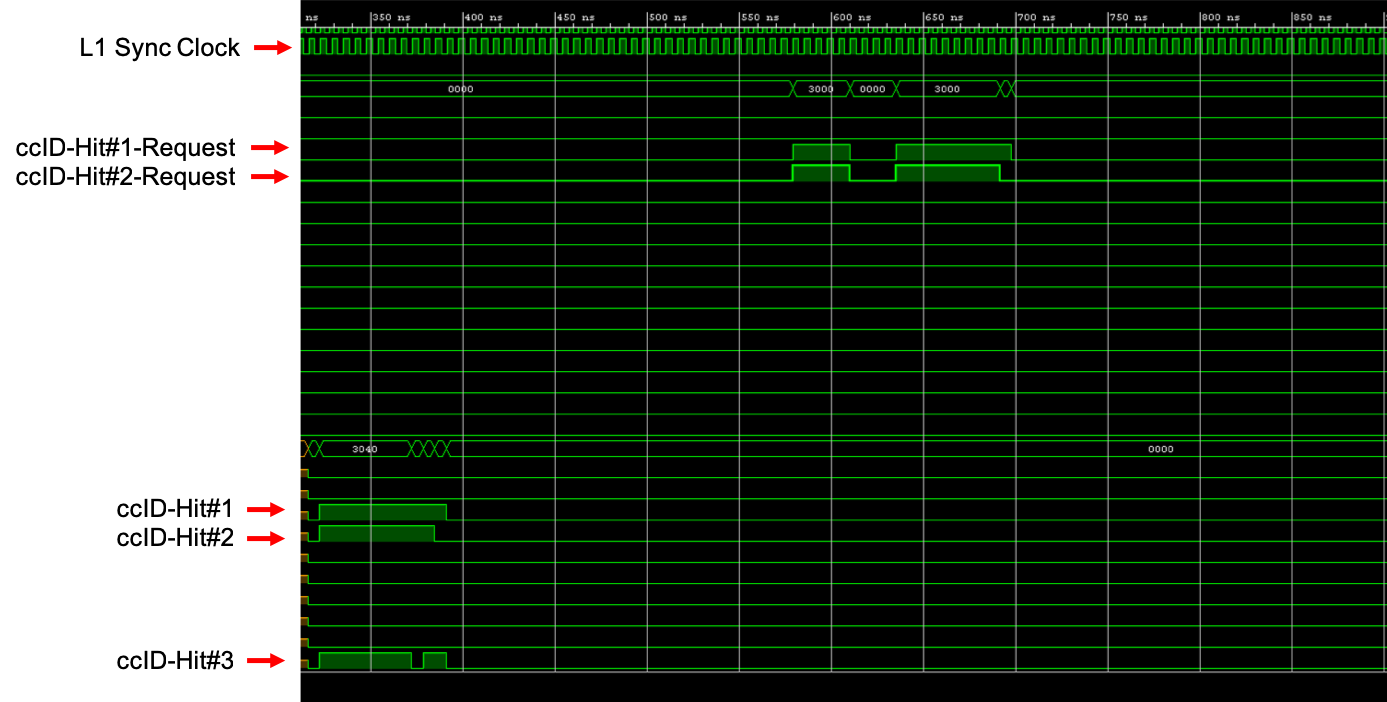}
\caption{Test bench signal logic, showing the L1 sync clock, 10\,bit ccID values for each hit that have been sent to the CTP, subsequently the CTP sends a full data request to the TC main board for hits \#1 and \#2.}
\label{fig:sim1}
\end{figure}

The test bench simulates the output of 3 sensors in separate super-modules. As previously mentioned the TC main board stores the data in dedicated RAM blocks in the FPGA and assigns each event a 10\,bit clock cycle ID (ccID) value. Figure\,\ref{fig:sim1} shows the 10\,bit ccID values that are sent to the CTP for coincidence checks. The first 2 hits, ccID-Hit\#1 (value\,$=$\,1023) and ccID-Hit\#2 (value\,$=$\,1022) pass the first coincidence check (being less that 2 clock cycles apart), the third hit, ccID-Hit\#3 (value\,$=$\,1019) does not pass this check and is discarded. As a result, the CTP only requests the full event data for hits \#1 and \#2 (ccID-Hit\#1-Request and ccID-Hit\#1-Request).\\

\begin{figure}[H] 
\centering
\includegraphics[width=0.99\textwidth]{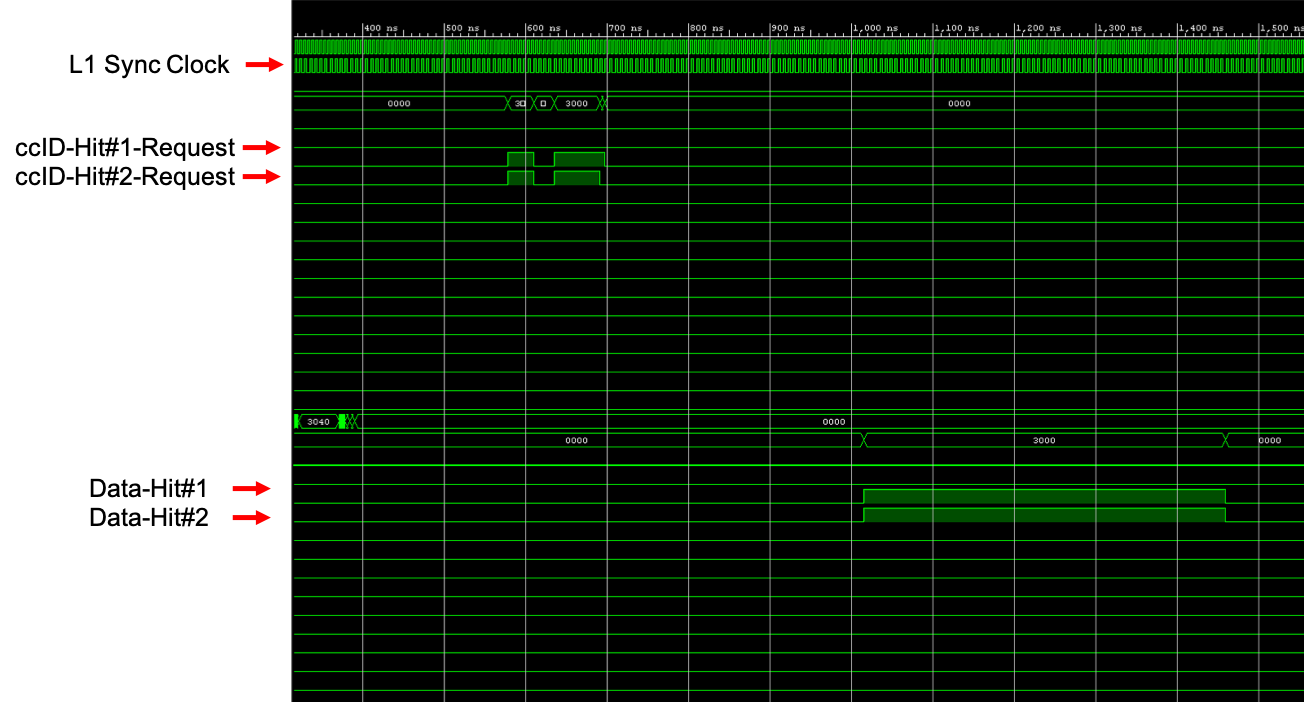}
\caption{Test bench signal logic, showing the L1 sync clock, CTP data request signals sent to the TC main board and the full event data being subsequently sent to the CTP.}
\label{fig:sim2}
\end{figure}

The TC main board sends the full event data for each super-module that passes the coincidence check to the CTP, see Figure\,\ref{fig:sim2} which shows the CTP event request, followed by the full event data for hits \#1 and \#2 that are sent to the CTP.\\

\begin{figure}[H] 
\centering
\includegraphics[width=0.99\textwidth]{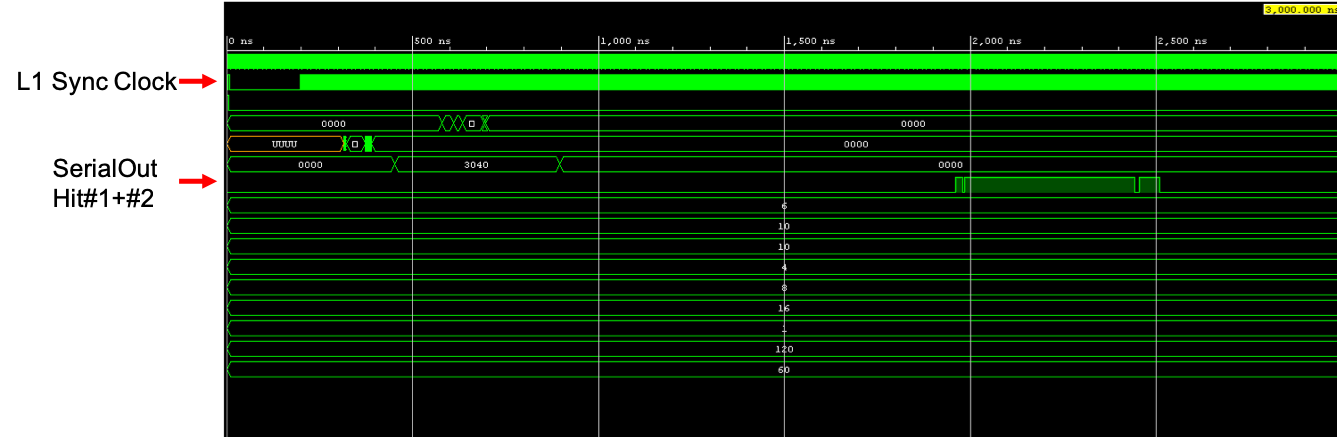}
\caption{Test bench signal logic, showing the L1 sync clock, and the serialised coincidence data sent from the CTP to the storage PC.}
\label{fig:sim3}
\end{figure}

The CTP combines the single hit event data, from the same coincidence event into a single serialised coincidence event. which is then sent to the storage PC for long term storage and processing. This is shown in Figure\,\ref{fig:sim3}. This hardware scheme is optional and the system can be use also in triggerless mode such that all the hits can be propagated downstream to the PC.\\

\subsection{Sensor Calibration}\label{subsec:calib}

The DAQ system is responsible for configuring and tuning the monolithic sensors. This configuration has been tested with the demonstrator prototype-0 chip \cite{ttpet:p0}. The system is able to set local and global register for the control of the discriminator threshold and  the frequency of the timing digital converter system.
A test setup consisting of the DAQ PC, TC main board, and prototype-0 monolithic sensor mounted on a configurable test PCB (Figure\,\ref{fig:TBV3}) was used to develop run control software to configure\,/\,setup multiple sensors and take event\,/\,noise data.\\
The prototype-0 PCB is embdedded with digital potentiometers for the control of the chip global threshold, the bias of the voltage controlled oscillator and the TDC one. These three additional chips are controlled via an SPI bus configured by the FPGA with internal devoted registers.
The prototype-0 test chip has 30 pixels in total. Each pixel is AC coupled to a fast Si-Ge BICMOS amplifier and discriminator, plus an 8 bit calibration DAC (TDAC). Every pixel is then multiplexed to a TDC, readout\,/\,programming logic and bias lines. The TDAC is used to control the in-pixel threshold in the discriminator to a certain value. When the signal is above the threshold the time is measured until the value drops below threshold. The tuning of the discriminator threshold is done in two steps. First, TDAC tune is performed to find global starting values for each pixel. Second, an in-pixel tuning is done by the TDAC based on the global values obtained during the first TDAD scan.\\

\begin{figure}[!hbt] 
\centering
\includegraphics[width=0.99\textwidth]{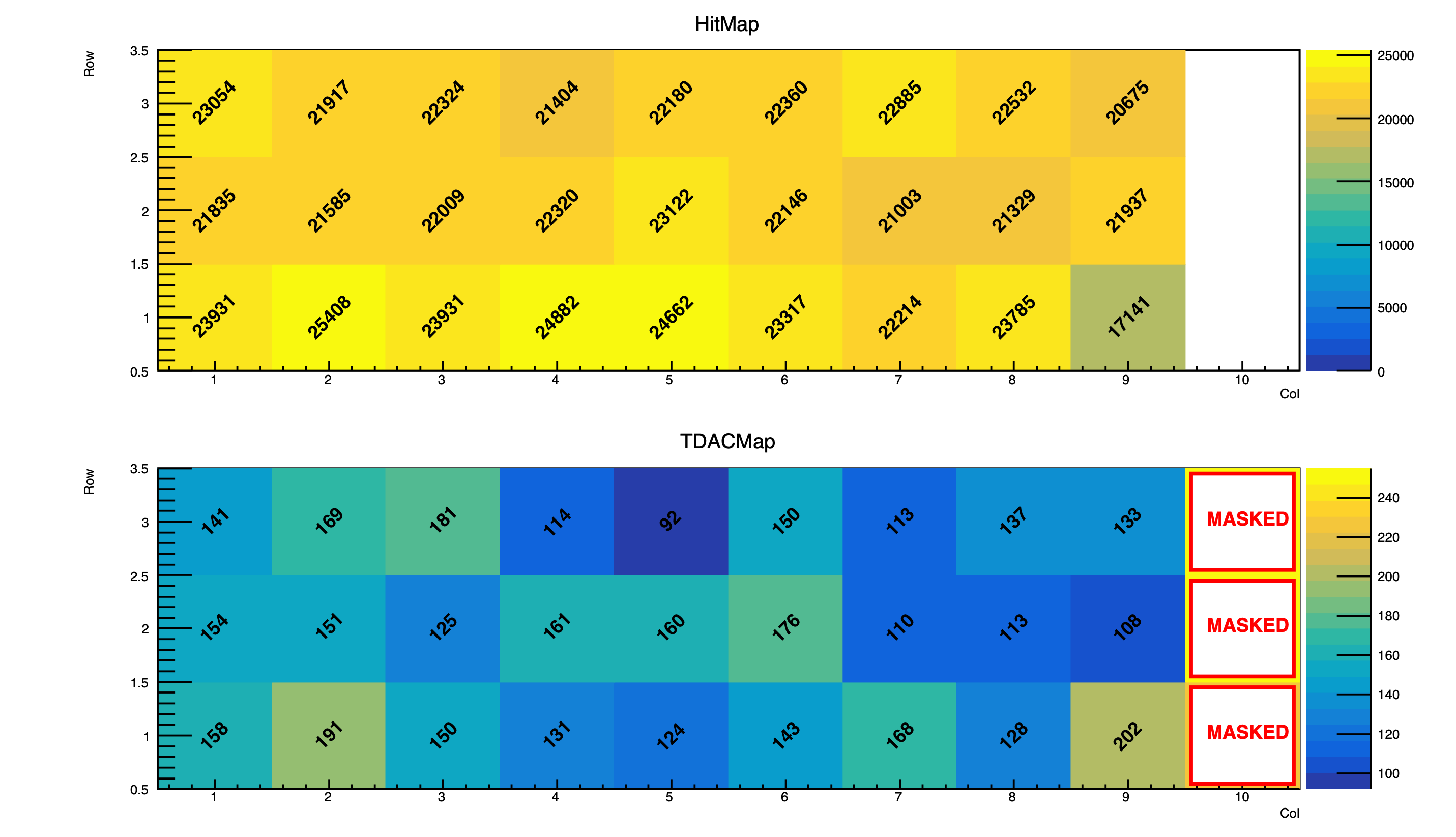}
\caption{2D Hitmap (top) and TDAC (bottom) tuning for each pixel in the prototype-0 monolithic sensor (3\,x\,10). The end column has been masked due to excessive noise.}
\label{fig:calib3}
\end{figure}

A TDAC tuning uses the bi-sectional method to iterate over many trigger threshold values, typically 6 seconds per setting is used to find the minimum noise free threshold setting for each pixel. This is done pixel by pixel, where all other pixels are masked (disabled) to exclude cross-talk. Figure\,\ref{fig:calib3} shows the results of a 2 stage TDAC tuning for a prototype-0 test chip. Top shows the number of events recorded by each pixel during the tuning. Bottom, shows the best TDAC values for each pixel.\\

\subsection{Clock Fan-out Characterisation}\label{subsec:jitter}

The quality of the L1 sync clock that is generated at the VLDB (section\,\ref{subsec:VLDB}) and distributed to each TC main board and subsequent super-modules is very important for an accurate time measurement. Since the L1 clock is used to generate the ccID values assigned to every recorded event and each sensors TDC uses the L1 clock when sampling the Time-over-Threshold (ToT) and Time-of-Arrival (ToA) information. Given the TDC binning of 30\,ps it was evaluated that the clock jitter should not overcome the level of 10\,ps to not affect the timing performance of the chip.

\begin{figure}[!hbt] 
\centering
\includegraphics[width=0.99\textwidth]{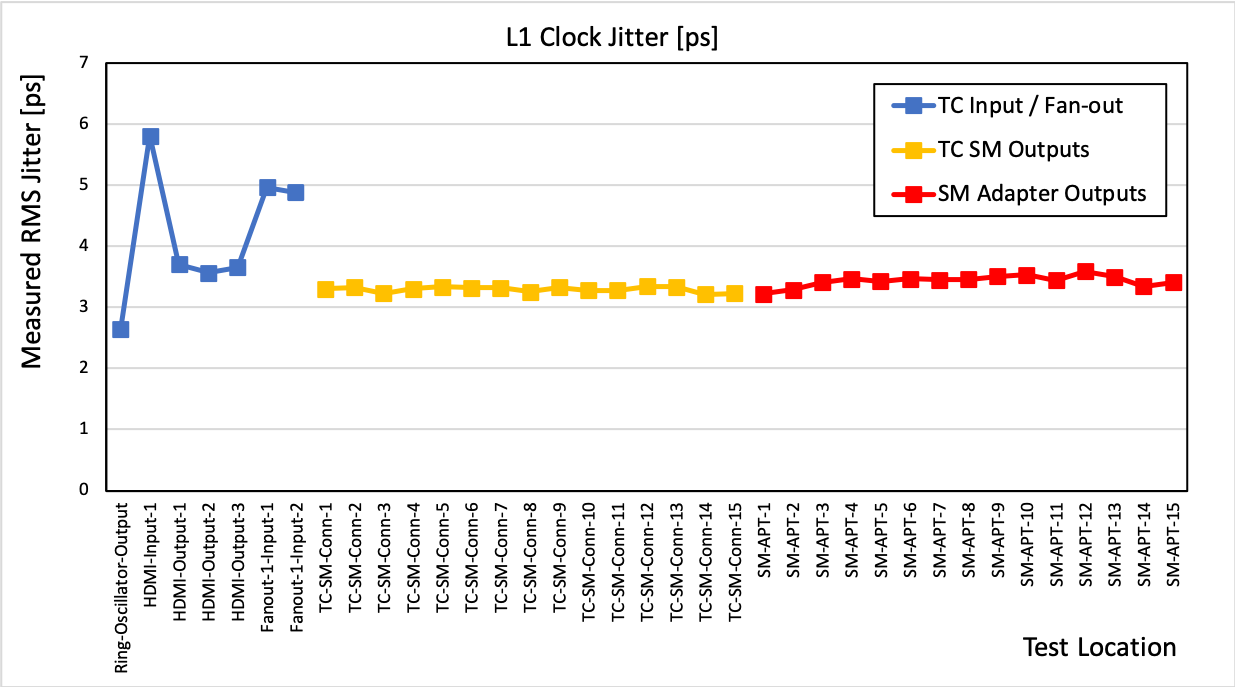}
\caption{RMS Jitter measurement of the L1 clock at various points along the clock distribution path on the TC main board and mock Super-Module (adapter pcbs). Blue line (TC Input\,/\,Fanout) are test points at the input of the TC and clock fanouts. Orange (TC SM Outputs) are test points at each of the 15 super-module output connectors. Red (SM Adapter Output) are test points using an external adpater pcbs connected to each super-module output connector.}
\label{fig:jitterplot}
\end{figure}

A 100\,MHz L1 clock is injected to the mini-HDMI eport on the TC main board using a differential ring-oscillator. The clock signal is sent to the on-board clock fan-outs, which sends the L1 clock to all 15 super-module output connectors on the TC main board. A Rhode \& Schwarz, RTO1044 digital oscilloscope with a 4\,GHz\,/\,20\,GSa/s bandwidth and differential probes are used to probe the jitter of the differential clock at various points along the clock distribution path on the TC main board. Figure\,\ref{fig:jitterplot} shows the measured jitter of the L1 differential clock at different test points. The blue points are measurements made at the clock generator, TC mini-HDMI input and on-board clock fan-outs. Each of the orange points corresponds to a separate super-module output connector. The red points are similar to the orange, with the exception of the use of an external pcb to minick the behaviour of a real super-module. In every measurement the clock distribution line is terminated at the super-module connector using a 100\,$\Omega$ resistor.\\

It can be seen that the measured jitter of the L0 clock distribution patgh on the TC main board is good, with an average RMS jitter of $\sim$\,3.5\,ps. The large jump in measured jitter at the input to the TC mini-HDMI eport and at the inputs to the 2 clock fan-outs is expected due to reduced shielding at the test points. The performance of the on-board clock fan-outs is excellent, providing the lowest average jitter signals with the exception of the clock generator (ring oscillator).\\

\begin{figure}[!hbt] 
\centering
\includegraphics[width=0.99\textwidth]{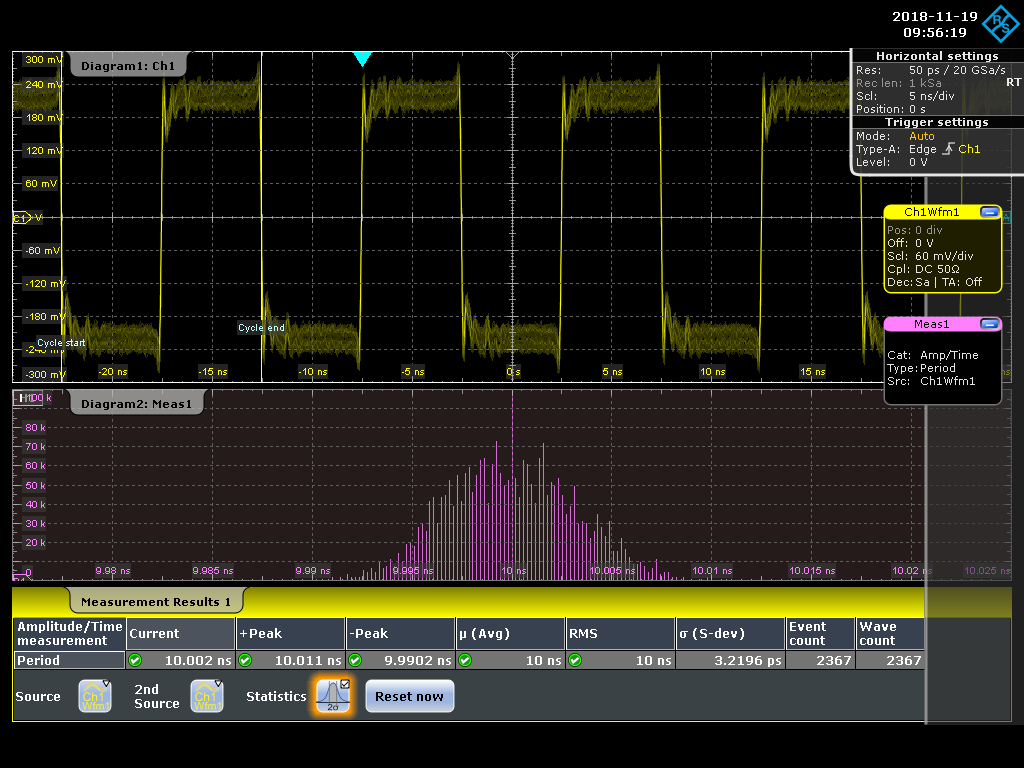}
\caption{Clock jitter measurement, showing the 100\,MHz clock in the top panel and their measured spread in the bottom panel. The average jitter has been calculated to be $\sim$\,3.2\,ps.}
\label{fig:jitt-func}
\end{figure}

The results shown in Figure\,\ref{fig:jitterplot} where calculated using the RMS-Jitter function of our digital oscilloscope. Figure\,\ref{fig:jitt-func} shows the measurement for a single Test Point ID. The top panel shows the L1 clock being measured, and the bottom panel shows the distribution as a function of time which is used to calculate the jitter (standard deviation, $\theta$ of distribution). The clock is measured using the rising edges of the clock period signals.\\

\subsection{Coincidence Measurement}\label{subsec:coin}

In order to test both the daq and monolithic sensors, coincidence measurements have been performed. As already mentioned in section\,\ref{subsec:trigger-overview}, a simplified version of the daq system, using only a single TC main board directly connected to the daq storage PC was used. This section will focus on the daq system and its ability to trigger on coincidence events rather than the performance of the prototype-0 sensor.

\begin{figure}[!hbt] 
\centering
\includegraphics[width=0.65\textwidth]{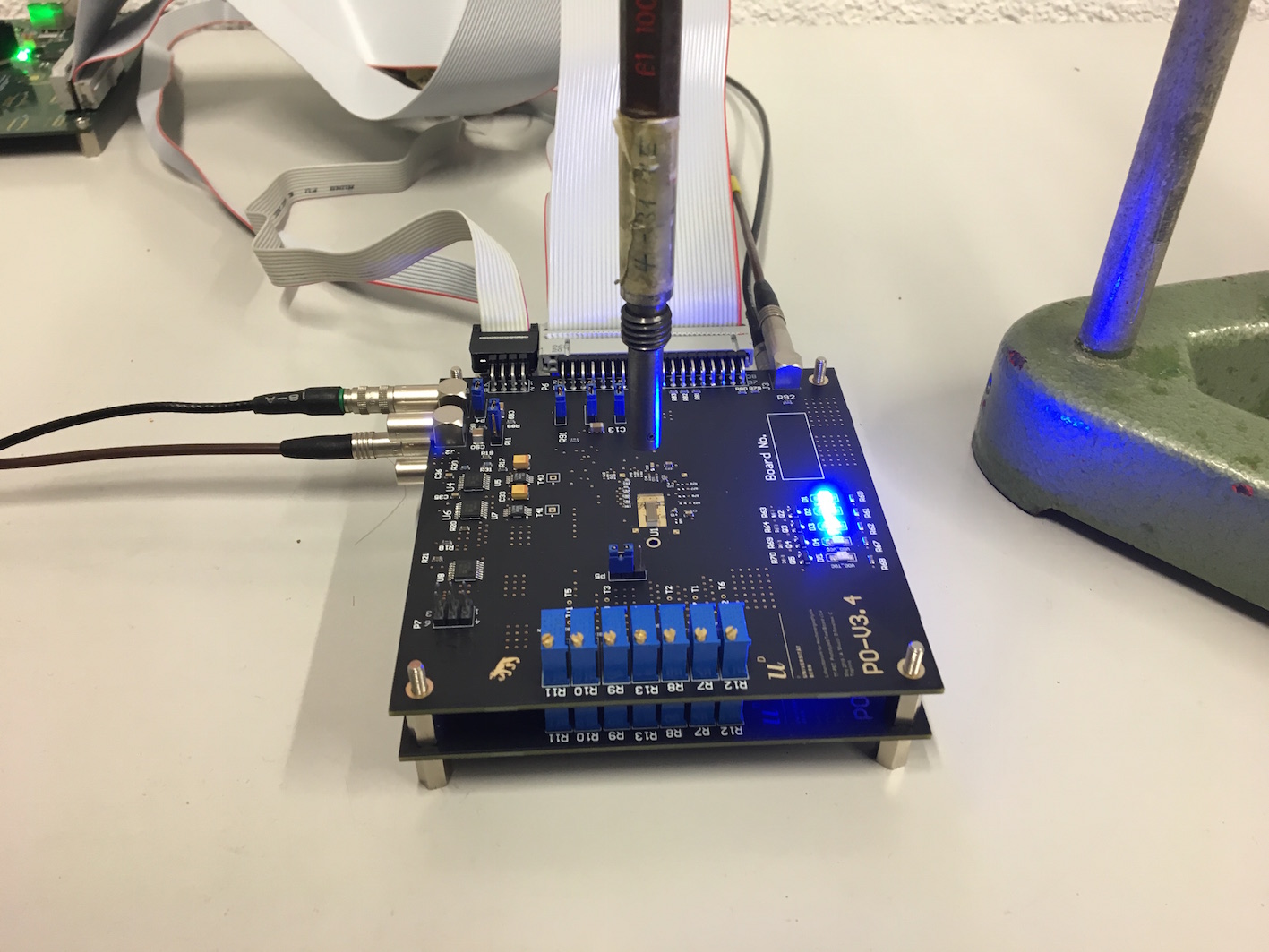}
\caption{Experimental setup for coincidence measurements with Sr-90 $\beta$-source. The electrons travel across the 2 sensors, in this way the coincidence can be detected. In order to minimise multiple scattering there is a hole in correspondance of the sensor loaded position.}
\label{fig:coin}
\end{figure}

Figure\,\ref{fig:coin} shows 2 prototype-0 sensors mounted on test PCB's, stacked on top of each other with a $^{90}$Sr $\beta^{-}$ radioactive source held directly above the center of the both pixel sensors. The top and bottom sensors will be referred to as chip0 and chip1 respectively.

\begin{figure}[!hbt] 
\centering
\subfloat[][]{\includegraphics[width=0.49\textwidth]{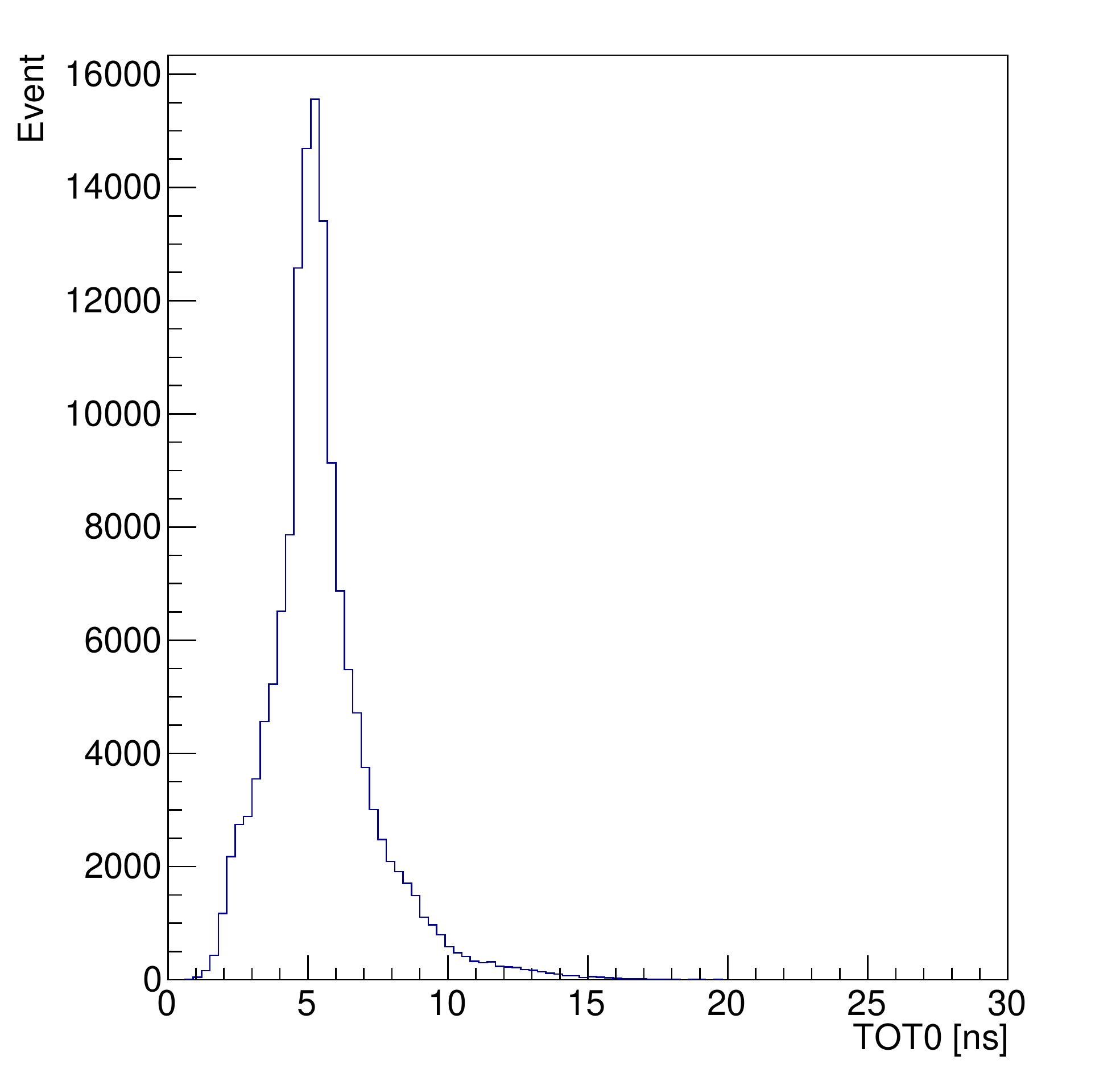}\label{fig:htot0} }
\subfloat[][]{\includegraphics[width=0.49\textwidth]{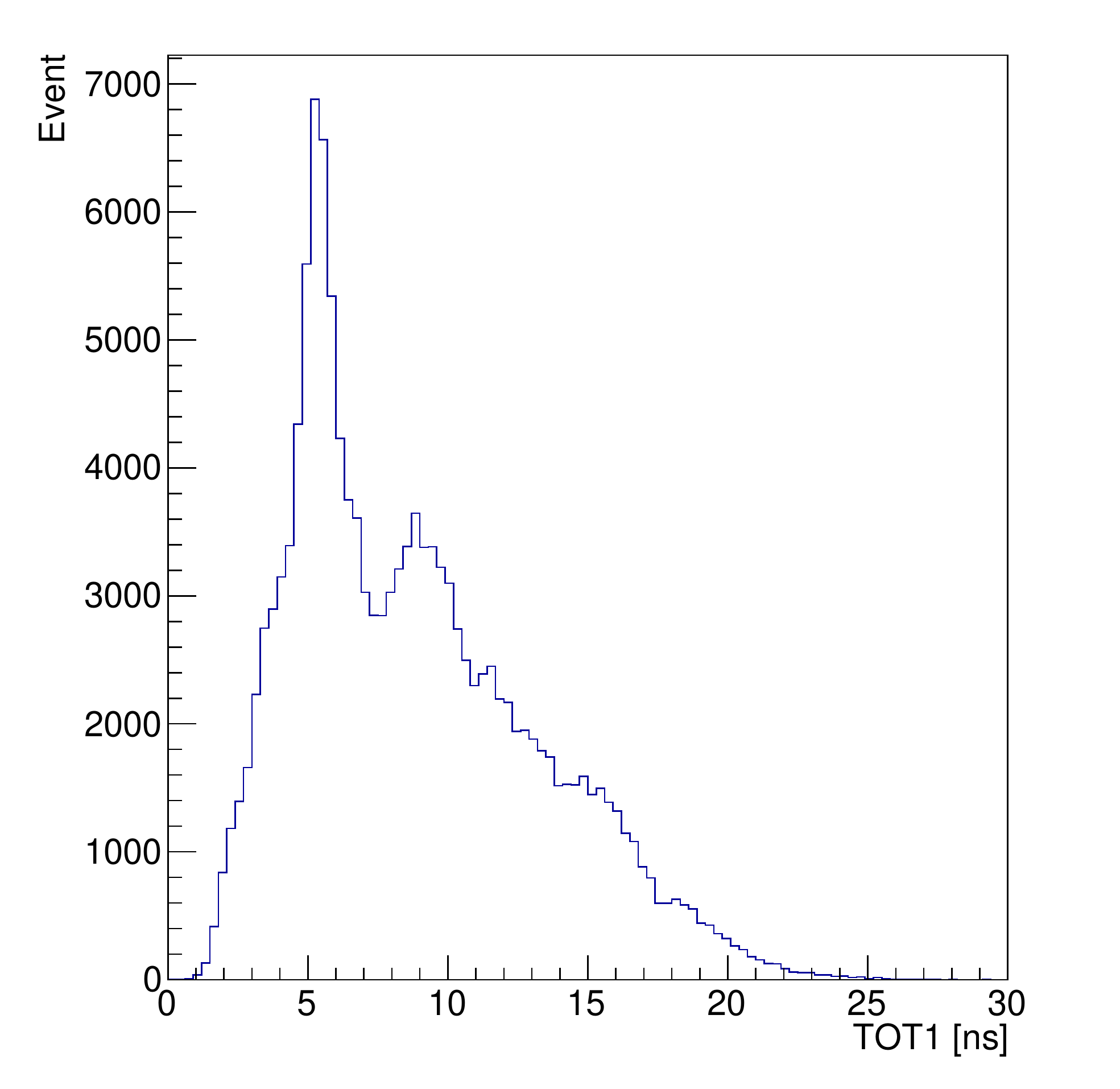}\label{fig:htot1} }
\caption{Time-over-Threshold distributions for \protect\subref{fig:htot0} chip0 (top sensor) and \protect\subref{fig:htot1} chip1 (bottom sensor). The higher ionisation in chip1 is due to the lower energy of the electrons in the second layer of the setup.}
\label{fig:coin2}
\end{figure}

\begin{figure}[!hbt] 
\centering
\subfloat[][]{\includegraphics[width=0.49\textwidth]{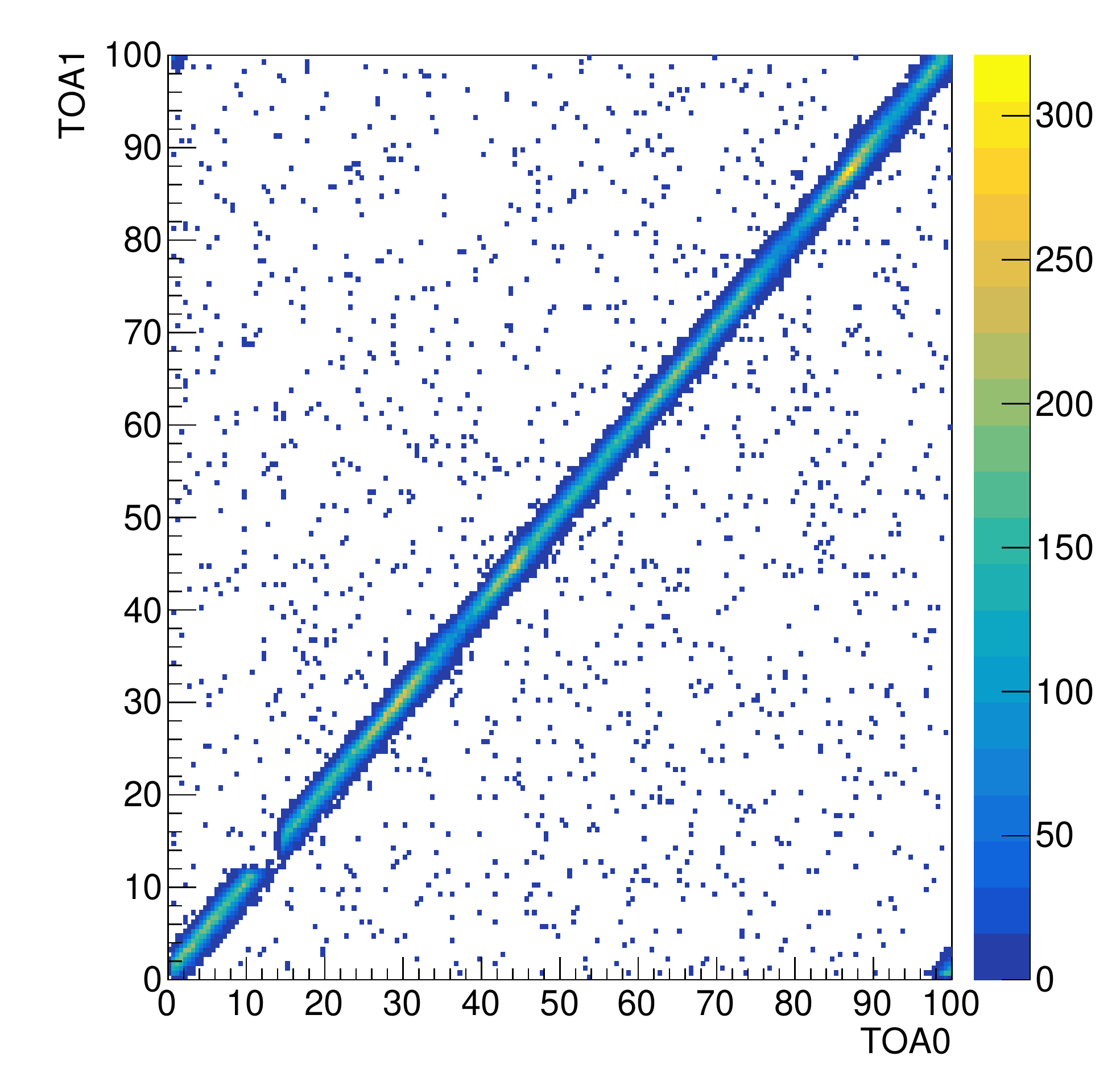}\label{fig:toa} }
\subfloat[][]{\includegraphics[width=0.49\textwidth]{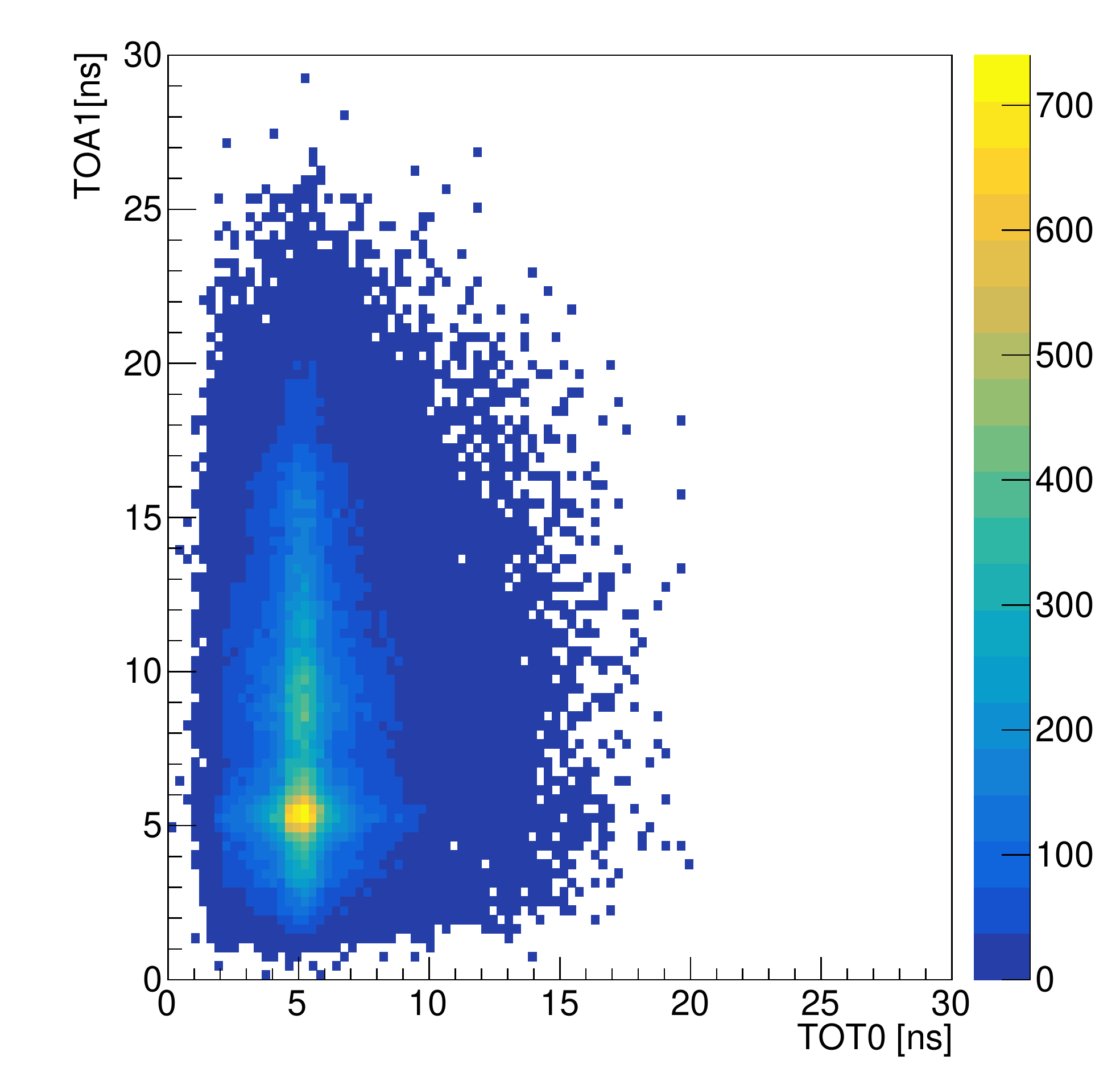}\label{fig:tot2d} }
\caption{Time-over-Threshold distributions for \protect\subref{fig:htot0} chip0 (top sensor) and \protect\subref{fig:htot1} chip1 (bottom sensor).}
\label{fig:coin3}
\end{figure}


\newpage
\section{Summary}\label{S:5}
 
A DAQ system was designed for the TT-PET project that allows to distribute low jitter clock to the scanner and read out and process the data up to a radiosotope activity of 50~MBq. In an emulation of the full readout chain the functionality of the elements and logic were established. 
A first implementation of a simplified prototype of the DAQ system was successfully used in a laboratory and test-beam setup with two prototype sensors, measuring coincidences of hits from a radioactive beta source and a beam of pions with 180\,GeV momentum.
The system-test demonstrate the capability of handling external triggers as well as the concept of self-triggering on coincidences. Data aggregation, event building and transfer to storage was as well proven.
Test bench measurements of the clock distribution show good performance with less than 4~ps jitter at the readout modules.
An updated version of the Tower Control board is in preparation to accommodate the readout of up to 60 sensing layers and cope with the design data rate.

\section{Acknowledgments}
The authors wish to thank the technical staff of the LHEP of the University of Bern and DPNC of the University of Geneva for their support. We thank the Swiss National Science Foundation, which supported this research with the SINERGIA grant CRSII2\_160808.




\newpage
\bibliographystyle{plain}
\bibliography{biblio.bib}







\end{document}